\documentstyle[12pt]{article}

\input epsf

\begin{document}

\begin{center}

{\Huge{Self-Screened Parton Cascades}}
\vskip.5truein

{\large Kari J. Eskola}

{Theory Division, CERN, CH-1211 Geneve23, Switzerland}
\bigskip

{\large Berndt M\"uller}

{Department of Physics, Duke University, Durham, NC 27708}
\bigskip

{\large Xin-Nian Wang}

{Nuclear Science Division, Lawrence Berkeley Laboratory}

{Berkeley, CA 94720}

\end{center}


\begin{abstract}
The high density of scattered partons predicted in
nuclear collisions at very high energy makes color screening effects
significant.  We explain how these screening mechanisms may suppress
nonperturbative, soft QCD processes, permitting a consistent
calculation of quark-gluon plasma formation within the framework of
perturbative QCD.  We present results of a model calculation of these
effects including predictions for the initial thermalized state for
heavy nuclei colliding at RHIC and LHC.
\end{abstract}

Most recent theoretical predictions for the initial conditions at
which a thermalized quark-gluon plasma will be produced at heavy ion
colliders are based on the concept of perturbative partonic cascades.
The parton cascade model \cite{GM95} starts from a relativistic transport
equation of the form
\begin{equation}
p^{\mu} {\partial\over\partial x^{\mu}} F_i(x,p) = C_i(x,p\vert F_k)
\quad i=q,g, \label{e1}
\end{equation}
where $F_i(x,p)$ denote the phase space distributions of quarks and
gluons.  The collision terms $C_i$ are obtained in the framework of
perturbative QCD from elementary $2\to2$ scattering amplitudes
allowing for additional initial- and final state radiation due to
scale evolution of the perturbative quanta.  To regulate infrared
divergences, the parton cascade model requires a momentum cut-off for
the $2\to2$ scattering amplitudes (usually $p_T^{\rm min}= 1.5 -2$
GeV/$c$) and a virtuality cut-off for time-like branchings $(\mu_0^2 =
0.5 - 1$ GeV$^2/c^2)$.

\begin{figure}
\def\epsfsize#1#2{.7#1}
\centerline{\epsfbox{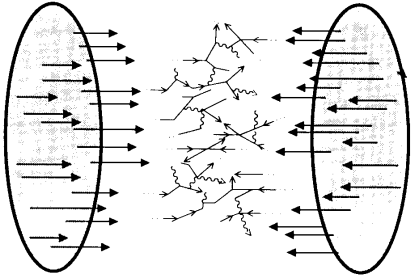}}
\caption{Schematic representation of a nuclear parton cascade.}
\end{figure}

Numerical simulations of such cascades for heavy nuclei provide a
scenario where a dense plasma of gluons and quarks develops in the
central rapidity region between the two colliding nuclei shortly after
the impact \cite{Gei95}.  Detailed studies \cite{EW94} indicate that 
the momentum spectrum of partons becomes isotropic and exponential, i.e.
practically thermal, at a time $\tau \approx 0.7 \Delta z$ in the rest
frame of a slab of width $\Delta z$ at central rapidity.  To permit 
a hydrodynamic description, the width of the slab should exceed the
mean free path of a parton.  Including color screening effects, one
finds that the mean free path of a gluon in a thermalized plasma is
$\lambda_f\approx (3\alpha_sT)^{-1}$ where $T$ is the thermal slope of
the parton spectrum.  For the predicted very high initial values of $T\;
(\ge 0.7$ GeV) one infers that a thermal hydrodynamic description
applies after $\tau_i\approx 0.3$ fm/$c$.

The high density of scattered partons in A+A collisions makes it
possible to replace the arbitrary infrared cut-off parameters
$p_T^{\rm min}$ and $\mu_0^2$ by dynamically calculated medium-induced
cut-offs \cite{Biro93}.  The dynamical screening of color forces 
eliminates the need for introduction of the momentum cut-off
$p_T^{\rm min}$, and the suppression of radiative processes provided by the
Landau-Pomeranchuk-Migdal effect makes the virtuality cut-off
$\mu_0^2$ unnecessary.  Note that the viability of this concept
crucially depends on the high parton density achieved in nuclear
collisions.  The dynamical cut-off parameters must lie in the range of 
applicability of perturbative QCD.  Since the density of initially 
scattered partons grows as $(A_1A_2)^{1/3}(\ln s)^2$, this condition 
requires both large nuclei and high collision energy.  The calculations 
indicate that this criterion will be met at RHIC and LHC but not at the 
presently accessible energies of the SPS and AGS. The framework is also not
applicable to $pp$ or $p\bar p$ collisions at current energies because
the parton density remains too low.

\begin{figure}
\def\epsfsize#1#2{.70#1}
\centerline{\epsfbox{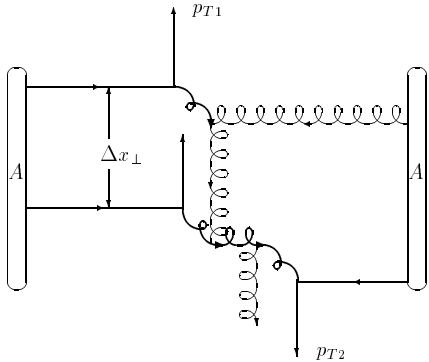}}
\caption{An example of the processes contributing to color screening.}
\end{figure}

The dynamic screening of parton cascades can be implemented as follows
\cite{EMW96}.  Consider an example of two hard processes as illustrated
in Fig. 2.  Let us assume that jets from the first hard scattering 
are produced at a large angle and carry a high transverse momentum
$p_{T1}$.  The interaction point is well localized transversely on a
distance of $\hbar/p_{T1}$.  As these two jets travel in the transverse
direction, they will experience secondary interactions, which can give 
rise to many nuclear effects of hard scatterings, e.g. energy loss and 
Cronin effects.  Here the interactions of the produced hard partons with 
the propagating partons originating from other perturbative scatterings 
nearby as shown in Fig. 2 are of interest.  

A semiclassical estimate of the screening requires that different
scattering events can be treated as incoherent.  This condition is
satisfied if the produced partons, which screen other softer
interactions, can be treated as on-shell particles.  This requires
that the transverse distance $\Delta x_{\perp}$ between the two
scatterings must be larger than the interaction range of the two hard
scatterings which are determined by the off-shellness of the exchanged
gluons.  If this condition is not satisfied, the propagating parton 
between two scatterings cannot be treated as real and consequently one 
cannot treat the multiple scatterings as incoherent.  We are thus led to
consider
\begin{equation}
\tau_f(p_T) = {a\hbar\over p_T} \label{e2}
\end{equation}
as the formation time of the produced partons in the mid-rapidity 
region from the hard or semihard scattering after which they can be 
treated as real (on-shell) particles and can screen other interactions with
smaller transverse momentum transfer.  The dimensionless coefficient $a$ 
of order unity parametrizes our uncertainty of the precise formation time.

In the framework of the inside-outside cascade, the incoming nuclei
being highly Lorentz-contracted, the primary semihard parton-parton 
collisions all start at $t=0$ and the evolving dense system at central 
rapidity is longitudinally boost invariant.  In the space-time evolution 
of the collision the partons with larger $p_T$ are produced earlier, as 
implied by (\ref{e2}).  These hard partons will then screen production 
of partons with smaller $p_T$ later in time and space.  Since, for fixed 
$p_T$, partons with larger rapidities form later in the chosen reference 
frame, only partons in the same rapidity range are relevant for screening.  
For the central region around $y=0$ we consider the screening effect of 
partons within a unit rapidity interval, $\Delta y =1$.

We now estimate the static electric screening mass
generated by the produced minijets.  The number distribution of
minijets produced in an $AA$ collision at an impact parameter $b=0$
can be written as \cite{BMW92}
\begin{equation}
{dN_{AA}\over dp_T^2dy} = T_{AA} (b) {d\sigma_{\rm jet} \over
dp_T^2dy}, \label{e3}
\end{equation}
where $T_{AA}(b)$ is the nuclear overlap function and
\begin{equation}
{d\sigma_{\rm jet}\over dp_T^2dy} = K \sum_{ijk\ell=q,\bar q g} \int
dy_2x_1f_i(x_1,p_T^2)x_2f_j(x_2,p_T^2) {d\hat\sigma^{ij\to k\ell}
\over d\hat t} (\hat s,\hat t,\hat u) \label{e4}
\end{equation}
is the minijet cross section.  The hats refer to the kinematical variables
of the partonic sub-processes, $x_i$ is the momentum fraction of the 
initial state parton $i$, $p_T$ is the transverse momentum, and $y$ is 
the rapidity of the final state parton.  The $f_i$ are the parton 
distribution functions, and $K=2$ is a factor accounting for the 
contribution from higher-order terms in the cross section \cite{EKS89}.  
For the purpose of screening we treat all the minijets as gluons.  This 
should again be a good approximation, since gluons clearly dominate the
minijet production \cite{EKR94}.

To obtain an estimate of the average parton number density in the
central region at a given time $\tau_f(p_T)$, we divide (\ref{e3}) by 
the approximate volume $V=\pi R_A^2\Delta z\approx \pi R_A^2\tau_f\Delta
y$ of the produced system.  Then the static color screening mass
becomes \cite{BMW92}
\begin{equation}
\mu_D^2(p_T) \approx {3\alpha_s(p_T^2) \over R_A^2\tau_f(p_T)\Delta y}\;
2\arcsin [\tanh (\Delta y/2)] \int_{p_T}^{\infty} dk_T {dN_{AA}\over
dk_T^2 dy}\Big\vert_{y=0}, \label{e5}
\end{equation}
assuming that all the quanta with transverse momenta $k_T \ge p_T$ 
screen the formation of partons at transverse momenta $k_T <
p_T$.  Only the quanta within the rapidity window $\Delta y$ are 
assumed to contribute to the screening mass.

In order to estimate the effect of this screening on the parton
scatterings with smaller $p_T$, we use the computed electric screening
mass as a regulator for the divergent $\hat t$- and $\hat u$- channel
sub-processes.  We will simply make a replacement $\hat t(\hat u) \to
\hat t(\hat u) - \mu_D^2$ in the minijet cross sections used in
(\ref{e4}).  In this way, by feeding the $p_T$-dependent screening mass 
back into the equation that defines it, we obtain self-consistent 
equations for the screening mass and the differential minijet cross 
section.  These equations can be solved numerically by starting at a 
large $p_T$ with no screening and then integrating down to smaller $p_T$.

\begin{figure}
\def\epsfsize#1#2{.70#1}
\centerline{\epsfbox{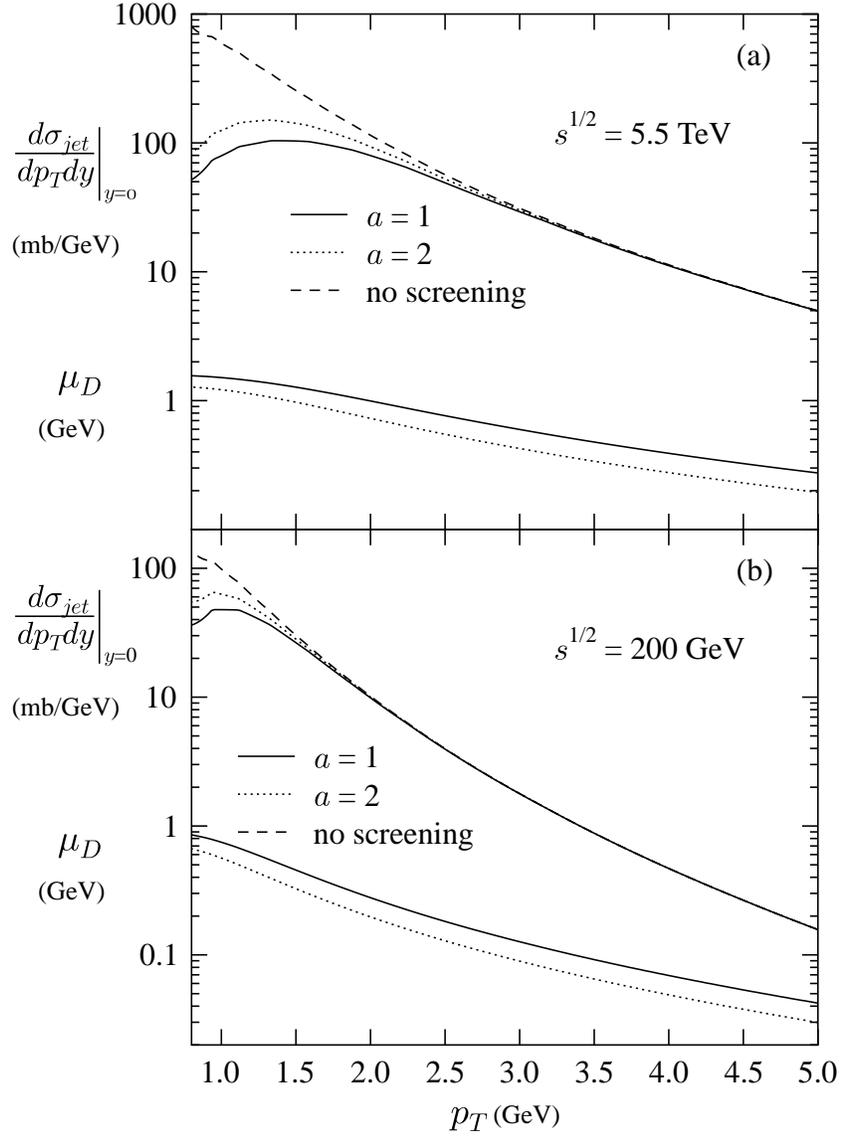}}
\caption{Differential minijet cross section at $y=0$ and screening mass 
$\mu_D$ as functions of transverse momentum $p_T$. (a) LHC energy,
(b) RHIC energy. Dashed line: without screening; solid and dotted lines: 
with screening.}
\end{figure}

In Fig. 3 we show the screening mass $\mu_D$ and the screened one-jet
cross section as functions of $p_T$.  In the upper panel the results
are shown for the LHC energy $\sqrt{s}=5.5$ TeV (per nucleon pair) and
in the lower panel for the RHIC energy $\sqrt{s} = 200$ GeV.
The jet cross sections are based on MRSA structure functions without
nuclear shadowing.  The
figure clearly supports our self-consistent picture of color screening:  
as the jet cross section grows, the parton medium becomes denser and
generates a large screening mass, slowing down the rise of the cross 
section towards smaller $p_T$.  In this way, the medium of produced 
minijets regulates the rapid growth of the jet cross section.  Finally, 
at $\mu_D\sim p_T$, the cross section saturates.  

To study the lack of sensitivity of the results to details of the 
uncertainty relation (\ref{e2}), we show curves corresponding to $a=1$ 
and $a=2$.  For A = 200 collisions at RHIC energy, the screening mass 
saturates at slightly below 1 GeV, and at 1.5 GeV for collisions at the 
LHC.  Both these values are comfortably within the
range of applicability of perturbative QCD, demonstrating that there
is no need for an artificial infrared cut-off.  The screening of
parton scattering by already scattered partons is analogous to the
interaction among ladders in the traditional picture of soft hadronic
interactions \cite{MS76}.  It would be interesting to rederive our
results from this alternative point of view.

\begin{figure}
\def\epsfsize#1#2{.70#1}
\centerline{\epsfbox{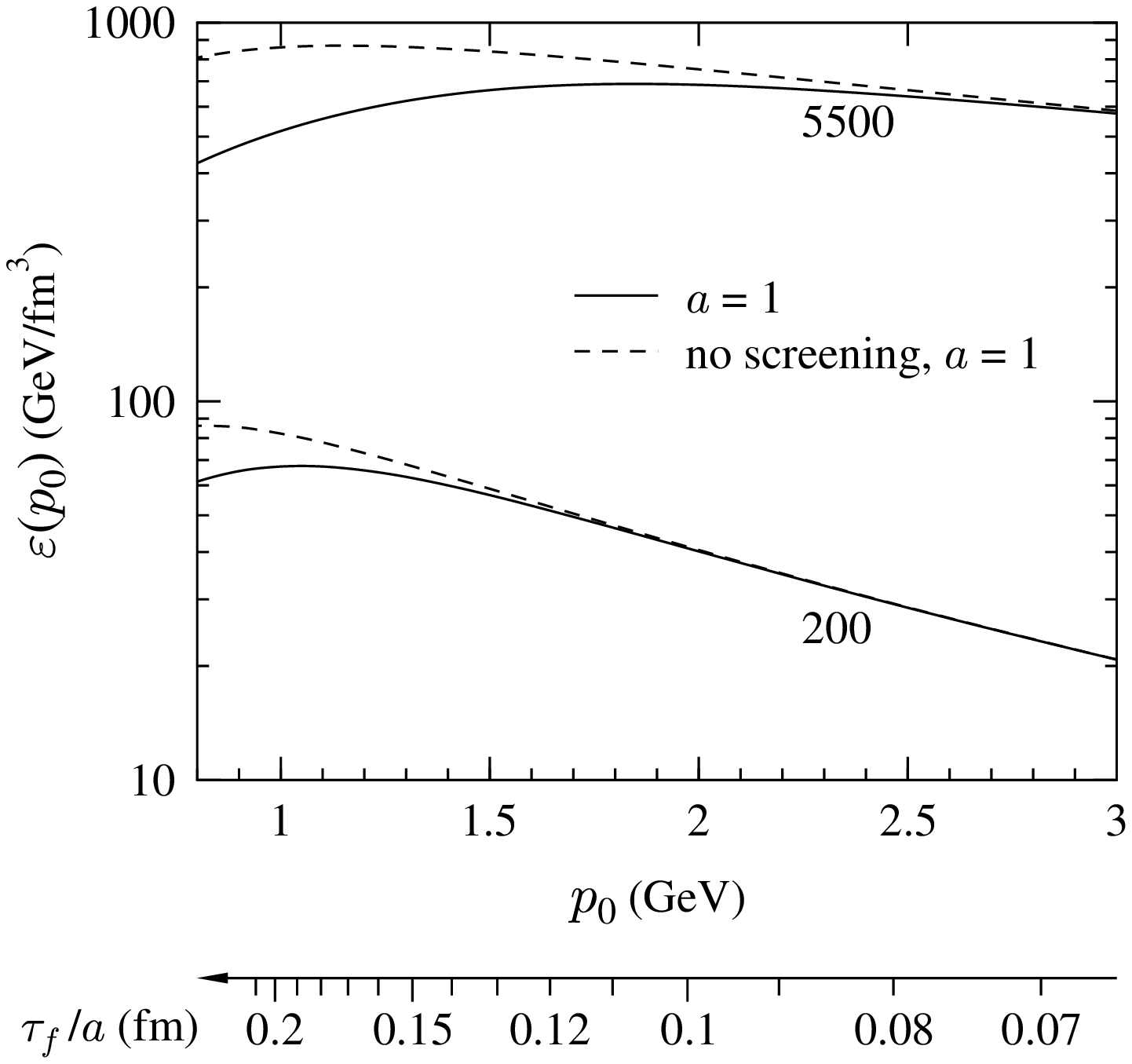}}
\caption{Transverse energy density $\epsilon$ of produced minijets as
a function of the lowest momentum transfer $p_0$ or the formation time
$\tau_f(p_0)$, respectively.  The solid and dashed lines show the
estimate with and without color screening.}
\end{figure}

\begin{table}
\begin{center}
\begin{tabular}{|l|r|r|} \hline
$\tau_i=0.25$ fm/$c$ &RHIC &LHC \\ \hline
$\epsilon_i$ (GeV/fm$^3$) &61.4 &425 \\
$T_i$ (GeV) &0.668 &1.02 \\
$\lambda_g^{(i)}$ &0.34 &0.43 \\ \hline
\end{tabular}
\end{center}
\caption{Initial conditions for the hydrodynamical expansion phase
at RHIC and LHC.  The initial time is taken as $\tau_i=0.25$ fm/$c$;
$\epsilon_i$ is the initial transverse energy density, and the 
effective number of flavors is assumed as $N_f=2.5$.} 
\end{table}

In order to study the further evolution of the dense parton plasma
created in the first generation of interactions, one can calculate the
energy density carried by the scattered partons.  The energy density
is obtained by dividing the total transverse energy produced by the
minijets with momentum transfer $p_T$ exceeding $p_0$ by the volume 
corresponding to the formation time $\tau_f(p_0)$:
\begin{equation}
\epsilon(p_0) = {E_T^{AA}(p_0)\over \pi R_A^2\; \tau_f(p_0)\Delta y}
\equiv \epsilon(\tau_f). \label{e6}
\end{equation}
The result is shown in Figure 4.  As a function of $\tau_f$ the energy
density first rises as more and more parton scatterings are completed,
but later starts to fall on account of the longitudinal expansion when
the saturation of minijet production due to color screening sets in.

Since earlier studies \cite{EW94,Biro93} have shown that the
conditions for a hydrodynamic description of the expansion are
satisfied at a time of order $\tau_i=0.25$ fm/$c$, for the energy
densities predicted by (\ref{e6}).  The full set of initial conditions
is listed in Table 1.  The initial temperature is predicted to be very
high in nuclear collisions both at RHIC and the LHC, but only about
one-third of the gluonic phase space is populated by the initial
parton interactions. We assumed here that the parton distributions become
isotropic due to free-streaming, and no additional transverse energy is
produced in the kinetic equilibration. (We emphasize that the assumptions
necessary for the conversion of our results into initial conditions for
the hydrodynamic evolution introduce considerable uncertainties into
the values listed in Table 1. These uncertainties could be eliminated
by a microscopic transport calculation of the kinetic equilibration
processes.)

Figure 5 shows the evolution of the temperature $T$, as well as the
gluon and quark phase space occupation ratios, $\lambda_g$ and
$\lambda_q$, as obtained from a longitudinal hydrodynamical expansion
with chemical equilibration \cite{Biro93}.  The equilibration only
accounts for the processes $gg\to ggg$ and $gg\to q\bar q$; it may
proceed faster if more complex reactions are also included
\cite{XS94}.  We have assumed that $\lambda_q^{(i)} = {1\over
5}\lambda_g^{(i)}$.  The evolution is stopped when the energy density
reaches 1.6 GeV/fm$^3$, where the transition to a mixed phase is
assumed to occur.  The lifetime of the pure plasma is found to be about 
4 fm/$c$ at RHIC and 18 fm/$c$ at the LHC.  For such a long life-time 
transverse expansion is expected to significantly reduce the plasma
life-time at LHC energies and to produce large collective transverse 
flow \cite{Sri96}.

Although many quantitative issues need to be resolved (e.g. the
influence of shadowing, the precise formation time, the correct value
for $\Delta y$) a well-defined picture of a parton cascade in nuclear
collisions, which screens its own infrared divergences, is emerging.
The screening mass $\mu_D(p_T)$ sets a scale which permits a
perturbative description of QCD interactions even in the limit
$p_T\to0$ as the parton density becomes high.  This concept is akin to
the picture of random classical color fields proposed in \cite{MV94}
for the small-$x$ gluon structure of large nuclei.  It is quite likely
that the two approaches can be connected.

\begin{figure}
\def\epsfsize#1#2{.70#1}
\centerline{\epsfbox{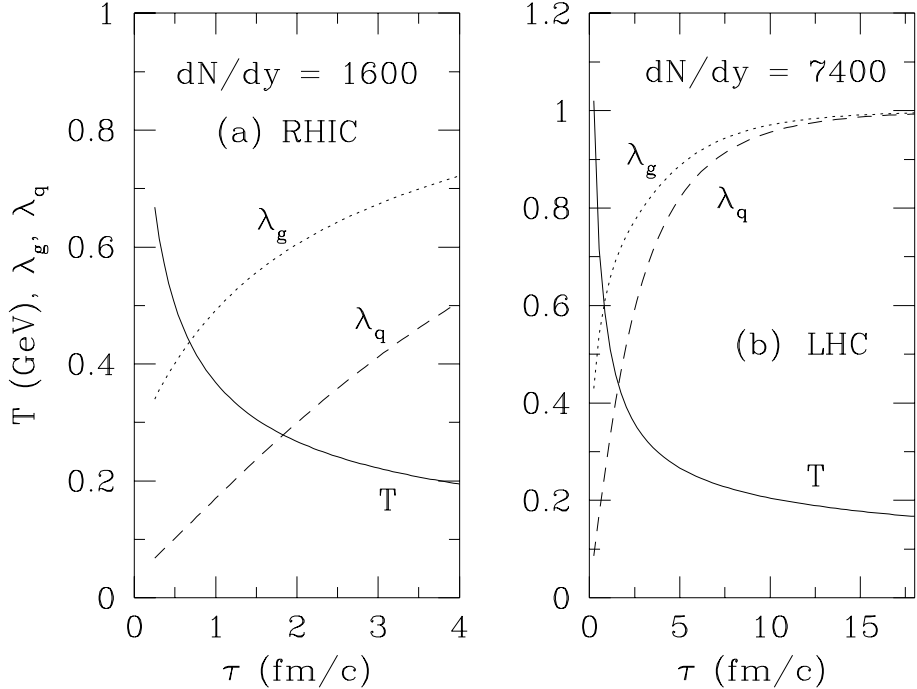}}
\caption{Evolution of the temperature $T$ and parton saturation
factors $\lambda_g,\lambda_q$ for the initial conditions given in
Table 1 in the longitudinal expansion model.}
\end{figure}

\subsection*{Acknowledgements}

This work was supported in part by the U.S. Department of Energy
(grants DE-FG02-96ER40945 and DE-AC03-76SF00098).

\end{document}